\documentclass[preprint,12pt]{aastex}
\usepackage{amsmath} \usepackage{amsfonts} \usepackage{amssymb}  \usepackage{apjfonts}\usepackage{color}

\pdfoutput = 1

\shorttitle{Transiting Exoplanets}
\shortauthors{Caitlin A. Griffith}

\begin{document}

\title{Disentangling degenerate solutions from primary transit and secondary eclipse spectroscopy of exoplanets}

\author{Caitlin A. Griffith\altaffilmark{1}}
\altaffiltext{1}{Univ. of Arizona, Dept. of Planetary Sciences, LPL, 1629 E. Univ. Blvd, Tucson, AZ 85721--0092; griffith@lpl.arizona.edu}

\begin{abstract}   
Infrared transmission and emission spectroscopy of exoplanets, recorded from primary transit and secondary eclipse measurements, indicate the presence of the most abundant carbon and oxygen molecular species (H$_2$O, CH$_4$, CO, and CO$_2$) in a few exoplanets. However, efforts to constrain the molecular abundances to within several orders of magnitude are thwarted by the broad range of degenerate solutions that fit the data.    Here we explore, with radiative transfer models and analytical approximations, the nature of the degenerate solution sets resulting from the sparse measurements of ``hot Jupiter'' exoplanets.  As demonstrated with simple analytical expressions, primary transit measurements probe roughly 4 atmospheric scale heights at each wavelength band. Derived mixing ratios from these data are highly sensitive to errors in the radius in planet (at a reference pressure), which are approximately a few percent.  For example, an uncertainty of 1\% in the radius of  a 1000 K  and H$_2$-based exoplanet with Jupiter's radius and mass, causes an uncertainty of a factor of $\sim$100-10000 in the derived gas mixing ratios. The degree of sensitivity depends on how the line strength increases with the optical depth (i.e. the curve of growth) and the atmospheric scale height. Temperature degeneracies in the solutions of the primary transit data, which manifest their effects through scale height and absorption coefficients, are smaller.  We argue that these challenges can be partially surmounted by a combination of selected wavelength sampling of optical and IR measurements and, when possible, the joint analysis of transit and secondary eclipse data of exoplanets. However, additional work is needed to constrain other effects, such as those due to planetary clouds and star spots.  Given the current range of open questions in the field,  both observations and theory, there is a need for detailed measurements with space-based large mirror platforms (e.g. JWST) and smaller broad survey telescopes, as well as  ground-based efforts.
\end{abstract}   

%\center{Accepted for publication in the Philosophical Transactions of the Royal Society A}

\section{Introduction}
\label{sc:intro}  

Over 800 planets have been detected outside the 
Solar System.  Current statistics show that planets are common; data 
from the Kepler Mission indicate that 
more than half of all stars have planets \citep{Fressin13}. Planets range in size, encompassing Earth to Jupiter diameters.  Most of these planets are smaller than Uranus; and terrestrial-sized planets are predicted to orbit one sixth of all stars \citep{Fressin13}.  However the majority of spectroscopically  
measured planets are predominantly Jovian-sized and  
hot ($T_{eff}$$\sim$700-3000 K).  It is largely, although not exclusively,  through
studies of these ``hot Jupiters and Neptunes'' as well as hot 
($T_{eff}$$\sim$500 K) ``super-Earths''  that 
techniques for measuring and retrieving 
compositional and structural information are currently tested. 
These techniques open the field of planetary sciences to 
the larger questions about planets related to their diversity,  
the physical processes that control their characteristics in different 
stellar environments, and the uniqueness of the Solar System and life.  

Studies of exoplanetary atmospheres began with observations of transiting 
planets (now over 230 of them), particularly the two bright exoplanet systems HD 209458b and HD 189733b.   
The first detection of a planetary atmosphere came from the analysis of optical spectra of 
HD 209458b as it passed in front of its host star \citep{Charbonneau00}. 
The planet's sodium resonance doublet at 589.3 nm was 
revealed from the attenuation of the star's light through 
the planet's atmosphere at the limb. Primary transit measurements of HD 189733b at near-IR 
wavelengths displayed features indicative of the presence of water, consistent 
with its expected large abundance and dominant role in the spectra of hot Jupiter exoplanets  \citep{Tinetti07,Grillmair08}. 
Thermal emission was first detected from measurements of HD 209458b's
passage behind its star during secondary eclipse. The planet 
plus star's 4.5 $\mu$m \citep{Charbonneau05} and 8 $\mu$m \citep{Deming05} 
emissions were compared to those of the star alone, yielding a
brightness temperature close to the current excepted value 
of 1130 K \citep{Fortney05}. 
Observations of exoplanetary transmission and emission spectra 
recorded during the primary transit and secondary eclipse, complement each other, as the former is 
sensitive to the planet's composition and not so much to the thermal profile, and the latter is 
sensitive to both of these planetary characteristics.  

%\begin{figure}
%\centering
%\rotatebox{0} {\includegraphics[type=eps,ext=.eps,read=.eps,width=3.4in] {figs/Exoplanetary-transits}}
%\caption{Measurements of transiting planets.} 
%\end{figure}

Currently, photometry and spectroscopy of transiting exoplanets indicates the 
presence of water, methane, carbon monoxide and carbon dioxide in a number of extrasolar planets, (e.g.  Tinetti {\it et al.} 2007, Grillmair {\it et al.} 2008, Madhusudhan \& Seager 2009, Swain {\it et al.} 2008, Swain {\it et al.} 2009b, Snellen {\it et al.} 2010).  
Observations at different points in an exoplanet's orbit also reveal variations in the planet's
the temperature field with longitude, which indicate the planet's dynamical redistribution of
heat, (e.g. Harrington {\it et al.} 2006, Knutson {\it et al.} 2007, Crossfield {\it et al.} 2010, Knutson {\it et al.} 2012). Yet even for the brightest systems, 
molecular abundances are constrained only to within
3-5 orders of magnitude and temperatures as a function of 
pressure constrained to roughly 300 K, (e.g. Madhusudhan \& Seager 2009, Lee {\it et al.} 2012, Line {\it et al.} 2012).  
The lack of precision in the derived characteristics 
results in part from systematic errors and noise commensurate with the small signal of the planet, 
and for certain systems from the host star's variability.  Yet many of the uncertainties  in the
derived composition and thermal structure of exoplanets, at present, 
stem from the range of models that fit the data, (e.g. Swain {\it et al.} 2009a, Madhusudhan \& Seager 2009, Madhusudhan {\it et al.} 2011, Benneke \& Seager 2012, Lee {\it et al.} 2012, Line {\it et al.} 2012, Griffith {\it et al.} 2013).  

Emission spectra, recorded during the 
secondary eclipse, can be interpreted with a range of temperature an composition profiles. 
Ultimately, if in local thermodynamic equilibrium (LTE),  
the absorbing gases must have abundances that cause emission from pressure
levels where the temperature approximates the observed brightness temperature. This
dependency leads to a number of degenerate solutions where the derived gas abundances 
correlate with the derived temperature profiles \citep{Swain09b,Madhu09,Lee12,Line12}. 

Light curves recorded during the primary transit measure the ratio of the 
planet to star areas, revealing the transmission of the light
through the limb of the planet's atmosphere. These measurements probe the planet's
composition, clouds, and the surface 
pressure \citep{Seager00,Brown01,Benneke12}. The retrieved gas abundances depend on the atmospheric 
temperature through the occupation of states, doppler line broadening,  
and the atmospheric scale height $H = {{R_gT}/{\mu g}}$.\footnote{Here $R_g$ 
is the gas constant, $T$ the temperature, $\mu$ the mean molecular
weight and $g$ the gravity. The scale height is a measure of the e-folding height of the 
atmospheric pressure at constant temperature, i.e. the puffiness of the atmosphere.} 
The derived abundances are thus affected by the mean molecular weight 
of the atmosphere \citep{MillerRicci09}, as well as  the surface pressure and radius 
as a function of pressure \citep{Tinetti10,Benneke12}.  Particularly for the low spectral resolution
observations currently possible, these dependences lead to the derivation of  a range of temperature
and composition profiles. 

\begin{figure}
  \epsscale{0.8}
  \plotone{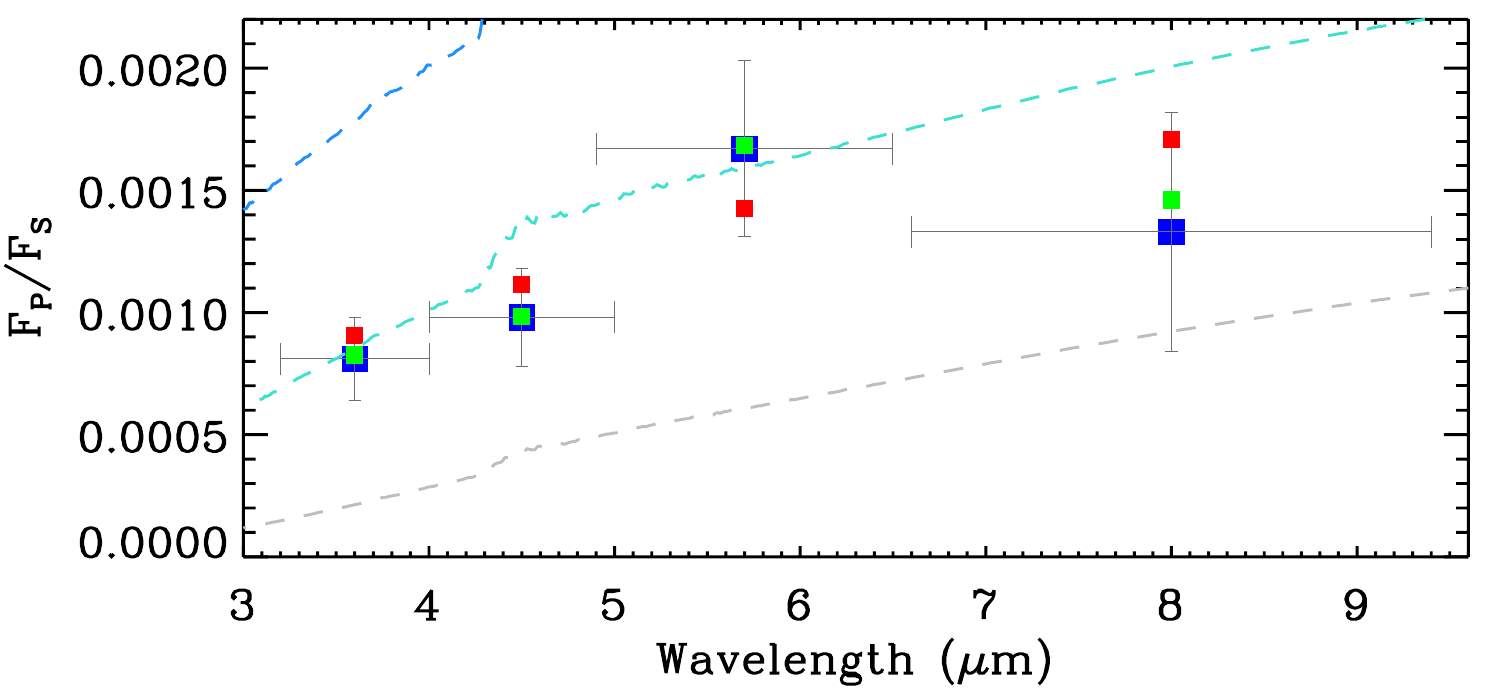}
% \plotone{mosesTP.eps}}
  \caption{SPITZER photometry measurements \citep{Machalek09} of the ratio of XO-2b's flux to that 
of its host star (blue squares), compared to a couple of models (red and green). Blackbody fluxes at 1000, 1500 K are shown in gray and cyan. Adopted from \citep{Griffith13}.}
  \label{fig:tp}
\end{figure}

%\begin{figure}
%\centering
%\rotatebox{0} {\includegraphics[width=3.8in] {XO2b-emission-final-RSb.pdf}}
%\caption{SPITZER photometry measurements \citep{Machalek09} of the ratio of XO-2b's flux to that 
%of its host star (blue squares), compared to a couple of models (red and green). Blackbody fluxes at 1000, 1500 K are shown in gray and cyan. Adopted from \citep{Griffith13}.} 
%\end{figure}

Here we explore the degenerate solutions of hot Jovian exoplanets, 
where the mean molecular weight is assumed to be dominated by 
hydrogen and helium, and any surface lies too deep to be detected. 
These assumptions pertain to all of the exoplanets extensively measured
to date, with the possible exception of the ``super-Earth'' exoplanet GJ1214b 
\citep{MillerRicci09}.
With this simplification, mainly the thermal profile, the abundances of 
the major carbon and oxygen molecules, and the planetary radius 
affect the analysis of infrared measurements of the primary transit and secondary eclipse. 
These characteristics are therefore the main observables. 
Other characteristics of the planet,  (e.g. clouds, 
minor species, and non-LTE emission), and the host 
star (e.g. star spots) can also play a significant role in the interpretation of 
exoplanetary data, and will be discussed briefly at the end of the paper. 
Yet currently, even ignoring these other characteristics,  there is too little 
data of high enough caliber, wavelength coverage and 
reproducibility to strongly constrain the characteristics of 
exoplanetary atmospheres,  largely because of
the degeneracies in the solution set.

In order to investigate the solution sets and 
information content of low resolution optical 
and near-IR measurements of exoplanetary atmospheres, 
we discuss the analysis of data of XO-2b, which has 
not been extensively measured. Thus the 
sources of the degeneracies are more obvious. 
The effects of the temperature 
and radius on the derived composition are found to 
depend on the planet's atmospheric scale height and can therefore 
be estimated analytically for any extrasolar planet. 
Analytical approximations, tested against a 
full set of radiative transfer models of data from the exoplanet XO-2b,
indicate that an uncertainty in the planet's radius of 1\% causes 
the gas mixing ratios derived from primary transit data to be uncertain
by several orders of magnitude, depending on the atmospheric 
scale height and the line regime. The analysis is not sensitive, separately, 
to uncertainties in the host star's radius, because primary eclipses 
measure the ratio of the planet to star radii. Thus the derived planet's 
radius scales to the assumed stellar radius, which must be indicated. 
In this study XO-2b's host star radius is assumed to be 0.964 $R_{Sun}$, \citep{Burke07},
where $R_{Sun}$ = 695500 km. 
This degeneracy and less so that 
introduced by uncertainties in the temperature, hamper 
the derivation of compositional information from primary transit data. 
In contrast, the largest correlations in the 
solution set of secondary transit data concern the thermal 
profile and composition; these degeneracies also yield abundances that range
several orders of magnitude.  

However for planets in which both primary and secondary 
eclipse data are possible, the joint analysis of both measurements 
decorrelates the radius and temperature and composition structures.
Here we assume that
the compositional and thermal differences between the sampled dayside and 
terminator atmospheres can be constrained, as
discussed further below. 
The analysis of XO-2b's sparse data indicates that the 
composition and thermal profile solution sets 
from primary and secondary eclipse data are 
significantly distinct that the combined analyses 
of these data yield stronger constraints on 
a planetary structure \citep{Griffith13}. 
Degenerate solution sets of primary and 
secondary eclipse are considered in detail to explore
the correlation between the parameters that lead to 
viable interpretations of current transit data.

%The question then arises as to how best to interpret the data available.  
%This paper explores the degeneracies in the solution sets of 
%primary transit and second eclipse observations 
%currently dominate investigations of transiting exoplanets. 
%We will first consider the degeneracies in models of XO-2b (Griffith et al. 2013), 
%and then attempt to understand these degeneracies with  
%simple analytical approximations.  

\section{LTE models of exoplanetary atmospheres} 

A number of different techniques are used to extract 
composition and temperature information from secondary 
eclipse data. All models start with basic assumptions 
regarding the number and identity of gases that affect 
the spectrum, and the temperature parameters
that characterize the thermal profile.  
The most basic extraction technique selects a phase 
space of all the potential mixing ratios and temperature 
profiles sampled at a fine enough grid to characterize 
the data within the errors \citep{Madhu09,Griffith13}. 
Another technique, the Markov chain Monte Carlo (MCMC)
method, similarly explores a phase 
space of potential solutions, calculates the spectrum, and
compares this to the data. However, this method does 
not directly calculate all of phase space. Instead it randomly jumps through 
while mostly accepting the jumps that improve the fit to the 
data, thereby converging to the solution sets more quickly 
\citep{Madhu11a,Benneke12}. Another approach, 
Optimal Maximization, inverts the data using an iterative
scheme to maximize the probability of attaining the best solutions to 
the data \citep{Rodgers00,Lee12,Line12}.  These models converge 
to a best solution. 
Here we adopt the first technique described 
above -- the brute force exploration of all of the defined phase space with millions of
models.  The advantage of this approach is that the analysis of both
primary transit and secondary eclipse can be explored
with the same parameter space or one altered to account for the differences 
in temperature and composition between the dayside and terminator 
atmospheres. After the spectrum of each model is computed, a principal components 
analysis (PCA) determines the strongest correlations between the
parameters in the solution set and thereby the linear combination of the
parameters that drive it. 

In this paper we consider, as an example, an analysis of the exoplanet XO-2b,
which is among the more extensively observed
``hot Jupiters'' in the sense that XO-2b is one of the few planets with a
Hubble Space Telescope (HST) primary transit spectrum  
from 1.2-1.8 $\mu$m \citep{Crouzet13} and secondary eclipse photometry 
from the space-based Spitzer telescope \citep{Machalek09}. 
%We discuss an analysis of  XO-2b's primary transit and 
%secondary eclipse data \citep{Griffith13} as an example of the 
%of the information attainable from transit spectroscopy.  
XO-2b orbits a K0V star 0.0369$\pm$0.002 AU 
away, which forms a binary system with a companion K0V star roughly 4600 AU from the host star \citep{Burke07}.   
The metallicity of this system is well established, with measurements of both stars indicating
the same composition within errors; the abundances of iron, 
nickel, carbon and oxygen are enhanced above solar \citep{Teske13a}.
With a 2.6 day period, XO-2b is a Jupiter-sized (0.996 $R_J$) planet, although slightly 
less massive (0.5 $M_J$) \citep{Burke07,Fernandez09}.

\subsection{Secondary Eclipse Measurements} 

Emission spectra of planetary atmospheres in LTE are controlled by 
the atmospheric temperature and opacity profiles. 
Likewise the thermal profile manifests the heating and cooling 
processes in the atmosphere. Hot Jupiter exoplanets, similar to the atmospheres of the giant
planets, are expected to have a convective region heated by the interior. 
However unlike the giant planets in the Solar System, for which  
the radiative-convective boundary lies at $\sim$1 bar, the highly irradiated 
hot Jupiters have boundaries at $\sim$100 bars.  The high optical depths in 
the region above this level in a hot Jupiter give rise to an isothermal region 
where radiative transfer proceeds by diffusion.  Consistent with model predictions \citep{Fortney05},  this stable region exists up to the pressure level ($\sim$1-10 bars) where the atmosphere starts to become optically thin. 
Above this level, the atmosphere cools radiatively, which causes the temperature to decrease 
with height, like the terrestrial troposphere. 
If higher up in the atmosphere there is a decrease in radiative cooling and/or 
an increase in heating, then the atmosphere's  
temperature increases, thereby creating a temperature inversion, similar to the 
stratosphere on Earth. 
For example, stratospheres form from the decrease 
in the pressure induced absorption with height, and 
through the presence of high altitude photochemical haze and other optical absorbers. 
These pressure-temperature regimes can 
be identified in theoretical models of the thermal profiles of the most extensively
observed exoplanets HD 209458b and HD 189458b (Fig. 2). 

\begin{figure}
\centering
\rotatebox{0} {\includegraphics[width=5.5in] {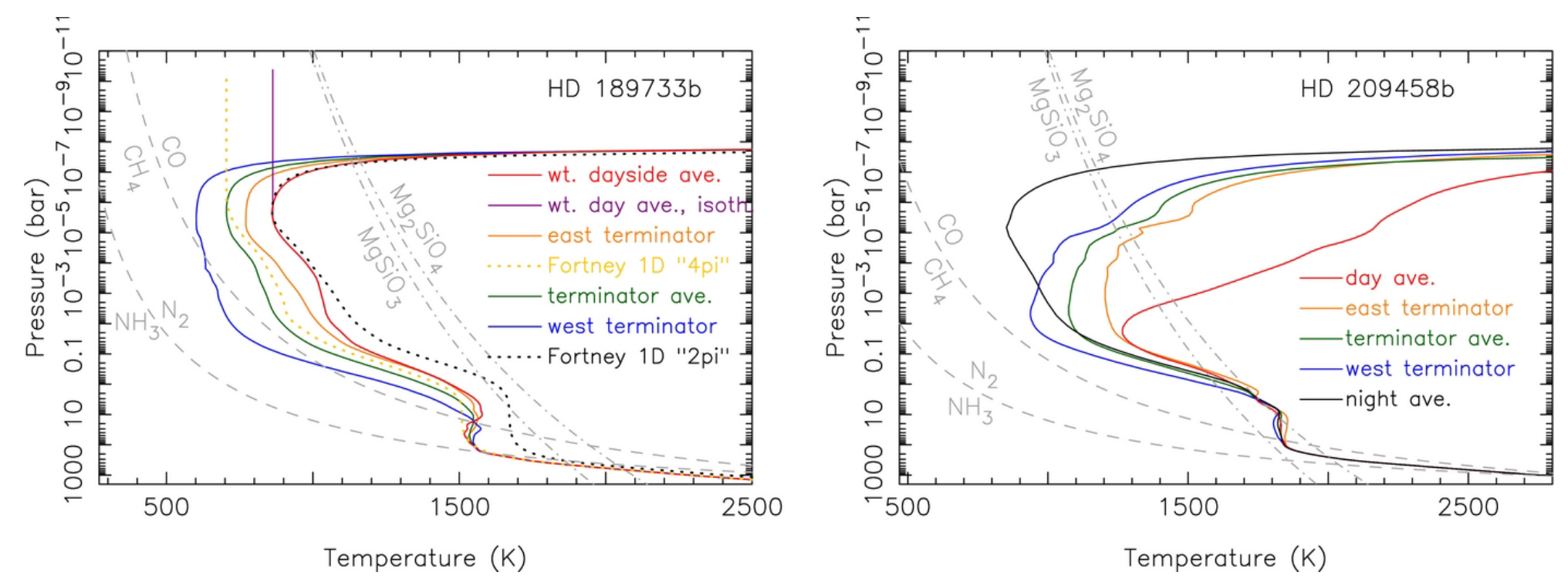}}
\caption{Temperature-pressure profiles of explanets HD 189733b and HD 209458b for the 
model atmospheres of Moses {\it et al.} 2011, based on the radiative transfer models of Fortney {\it et al.} 2006 and Fortney {\it et al.} 2010, 
and the GCM models of Showman {\it et al.} 2009.  Figure taken from Moses {\it et al.} 2011.}
\end{figure}

The presence of spectral features and their nature, i.e. whether they appear as 
an emission or absorption feature, depend on the slope of the section of the thermal 
profile being sampled.  If the temperature decreases with height, an
absorption feature causes a decrease in emission;  conversely if the temperature increases
with height; an absorption feature causes an increase in emission.  No features
are formed in a vertically isothermal region.   Measurements that sample 
both the stratosphere and troposphere of an atmosphere are difficult to interpret because 
an absorption feature can increase, decrease, or not affect the planet's 
outgoing intensity, depending on where it forms.   Most notably, 
interpretations of measurements of emission data of HD 209458b indicate 
either water features from the stratosphere \citep{Burrows07} or 
mainly methane absorption coupled with water
from the troposphere \citep{Swain09b}. The former interpretation \citep{Burrows07} however can not 
explain the Spitzer 24 $\mu$m measurement, and has most of the exoplanet's radiation 
emitted from the stratosphere.  
It turns out that an inverted water
spectrum can (with sparse data) 
resemble a methane spectrum over certain wavelength regions 
and at low spectral resolution.  However, in 
general most of the radiation from an explanet in the wavelength region 
of highest emission (i.e. near 3 $\mu$m for a $T_{eff}$=1000 K  planet)  
comes from the region where the temperature decreases with height 
due to radiative cooling. 

%\begin{table}
%\centering
%\caption{Model Parameters}
%\begin{tabular}{lcc}
%\hline
%Parameter     &    Range       &    Number     \\
%\hline
%[H$_2$O]       & 10$^{-3}$--10$^{-8}$ 	& 13       \cr  
%[CH$_4$]       & 10$^{-3}$--10$^{-8}$ 	& 13       \cr 
%[CO]               & 10$^{-3}$--10$^{-8}$ 	& 13       \cr 
%[CO$_2$]       & 10$^{-3}$--10$^{-8}$ 	& 13      \cr 
%$T_T$            &  600--1300 K                   & 8          \cr
%$T_I$             &  1300--3100 K                 & 24          \cr
%$P_T$            &  0.1--10$^{-5}$ bar       & 5         \cr
%$P_I$            &  1--10 bar                       & 3         \cr
%\hline  
%\end{tabular}  
%\end{table}

With the cautionary note that constraints on the composition and structure of an exoplanet depend on the spectral modulation of its features, the signal-to-noise, and the spectral resolution and coverage, here we examine, for the purpose of providing a simple example, an analysis of Hubble Space Telescope (HST) photometric measurements of XO-2b at 3.6, 4.5, 5.8 and 8.0 $\mu$m \citep{Machalek09}.  With such little data, 
only a simple atmospheric profile with no stratospheric inversion is considered here,  
which is consistent with theoretical calculations of the close 
analog system, HD 189458b \citep{Grillmair08,Madhu09,Lee12}. Models with 
temperature inversions are discussed in \citep{Griffith13}. 
The emission points are interpreted with a radiative transfer model that assumes 
LTE, and includes the opacity of H$_2$O, CO, CH$_4$, CO$_2$ 
and pressure-induced H$_2$.  As discussed in more detail in Griffith {\it et al.} 2013, 
line-by-line calculations of the HITEMP line parameters \citep{Rothman09}
are used to calculate k-coefficients of CO. For CH$_4$
and H$_2$O, k-coefficients were calculated from the
absorption coefficients derived by Freedman {\it et al.} (2008); 
absorption by CO$_2$ lines derive from line-by-line 
calculations of the CDSD data base \citep{Tashkun03}. 
However,  for CH$_4$, at 
wavelengths below 3 $\mu$m, we increase the methane absorption 
by a factor of 10 to fit the laboratory data at 2.2 and 1.7 $\mu$m \citep{Nassar03,Thievin08}.   

The temperature profile was parametrized 
with 4 variables in order to explore a range of 
profiles.\footnote{The thermal profile parameters are the temperatures of the troposphere, $T_T$, 
and deep isothermal atmosphere, $T_I$, and the pressures of the tropopause, $P_T$,  
and the top of the lower isothermal region, $P_I$, following \citep{Madhu09}.
The range of values considered are, respectively, 600-1300 K, 1300-3100 K, 0.1--10$^{-5}$ bar 
and 1--10 bar. In total 2880 profiles were calculated. } 
The mixing ratios are assumed constant, which is the expected profile for 
H$_2$O and CO, and a reasonable one for CH$_4$ and CO$_2$ \citep{Moses11},  since the data 
sample only a small pressure region (Fig. 3). Mixing ratios ranging from 
10$^{-3}$--10$^{-8}$ were considered for each molecule, except for CO$_2$ which 
ranges 10$^{-4}$--10$^{-8}$, because it is 
expected to be significantly less abundant \citep{Moses11}. Based on the planet's density, 
the atmosphere is taken to be mainly hydrogen and helium with 
Jovian mixing ratios. Radiative transfer calculations derive the flux 
at the 4 measured photometry points, 
integrated over each Spitzer filter bandpass for each model atmosphere.
Given the range and sampling of gas mixing ratios and the   
thermal profile parameters a total of 
almost 82 million non-inversion models were calculated and compared to the data. 
(The original study also includes temperature inversions \citep{Griffith13}.)
The goodness of fit 
is evaluated with a weighted mean square error, $\epsilon$, such that: 
$$
\epsilon = {1\over N} ~ \sum_{i=1}^N ~ ({{F_i^{model} - F_i^{obs}}\over{\sigma_i}})^2
$$
where $F_i^{obs}$ is the observed planet flux to 
star flux ratio at $i^{th}$ wavelength in the spectrum, 
${F_i^{model}}$ is the calculated value, and $\sigma_i$ is
the 1 sigma error in the measured value \citep{Madhu09}. 
However, since the errors do not necessarily follow a gaussian distribution, 
we keep all models that fit the data within the errors.

\begin{figure}
\centering
\rotatebox{0} {\includegraphics[width=4.5in] {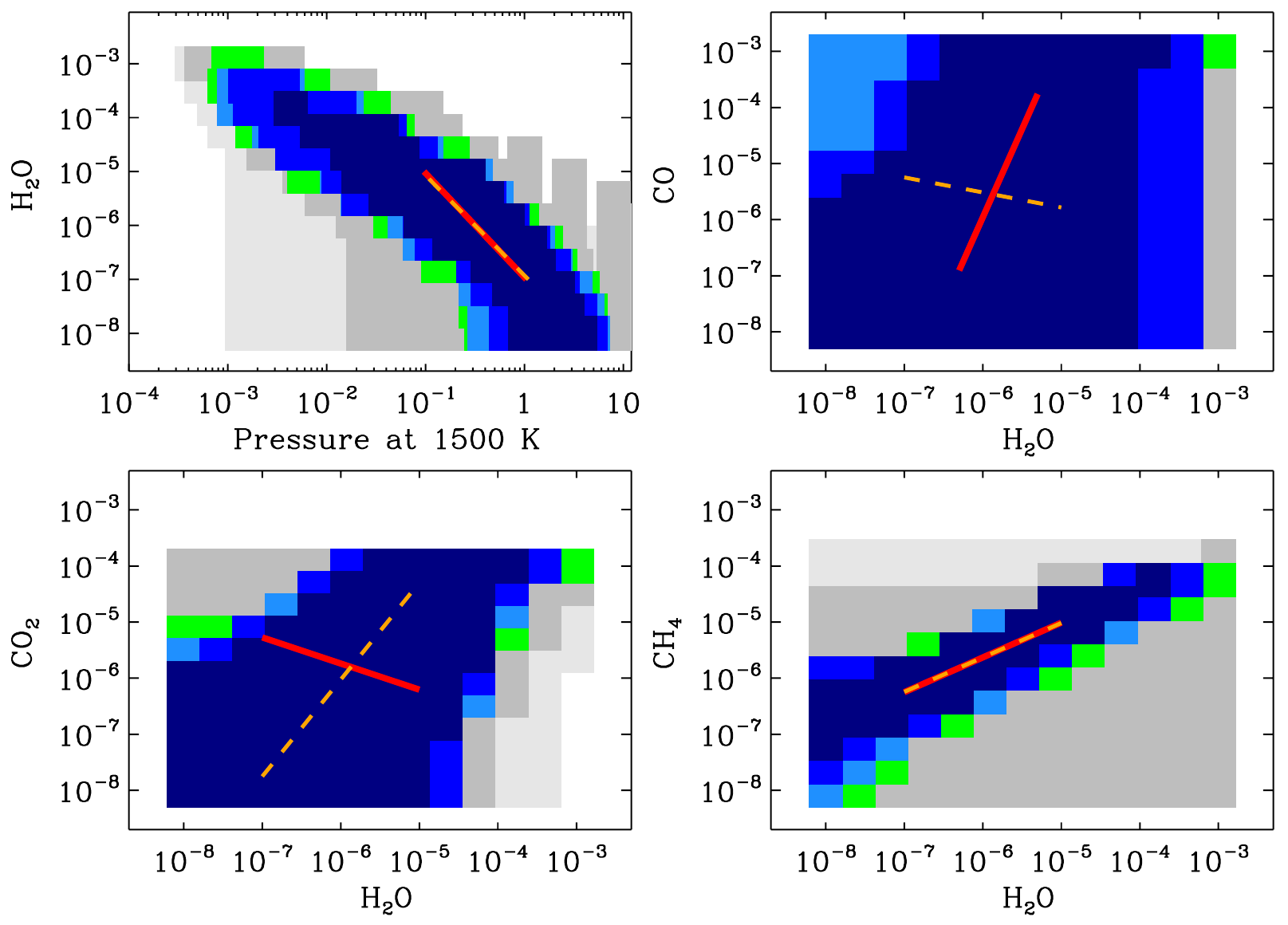}}
\caption{Models that match XO-2b's secondary 
eclipse data within the errors are shown for two model 
parameters in each panel. Colors represent the degree
to which the models match the data; navy blue, blue, light blue, green, and grey 
correspond to weighted arithmetical mean square errors increasing respectively from  0.2 to 0.5; light grey indicates
values above 0.5. Navy blue models most closely match the 
data. Red and orange lines mark the projection of the principal and secondary axes of the principal 
component analysis of the solution set as described in Griffith {\it et al.} 2013. Credit: Griffith {\it et al.} 2013.} 
\end{figure}

While none of the parameters are constrained, only 3\% of the explored phase space yielded 
models that fit all 4 emission points within the errors. In addition, the parameters of 
the successful models are highly correlated. As shown in figures 3--4, the abundance of water, which 
controls most of the opacity, tracks with the pressure level 
of the 1500 K atmospheric level. The abundance of methane also plays a large 
role in the models that fit the data, while CO and CO$_2$ similarly influence only the interpretation 
of the 4.5 $\mu$m measurement (Fig. 3).  These correlations in fact drive the population of the 
solution, as can be seen from the principal component and secondary component of the 
Principal Component Analysis (PCA) of solution set, which line up with the apparent correlations 
in the gas abundances (Fig. 3). Note that abundances of CO and CO$_2$ compete in
establishing the 4.5 $\mu$m measurement, and establish the secondary component correlation. 
The PCA analysis is discussed in more detail by Griffith {\it et al.} (2013). 

This solution is not surprising; essentially, the mixing ratios fit the data as long as the 
temperature profile is adjusted so that the outgoing intensity is equivalent to the measured brightness 
temperature of 1000--1600 K (Fig 1).   The other gases track 
loosely with water, with the exception of CO$_2$ and CO, which has 
little effect on the spectrum.  Similar trends are detected in 
the analyses of the HD 209458b \citep{Swain09b}. However with such sparse data, no constraints could be derived for individual parameters, i.e. those that 
define the thermal profile and the composition. For example, atmospheres with and without thermal inversions fit the data \citep{Griffith13}. 
%The solution set is somewhat straightforward, and dependent on the model assumptions. 
The addition of a temperature inversion of, for example, 800 K leads 
to additional peaks in the contribution functions and therefore separate sets of correlated solutions associated with each stratospheric profile \citep{Griffith13}.  

\begin{figure}
\centering
\rotatebox{0} {\includegraphics[width=4.5in] {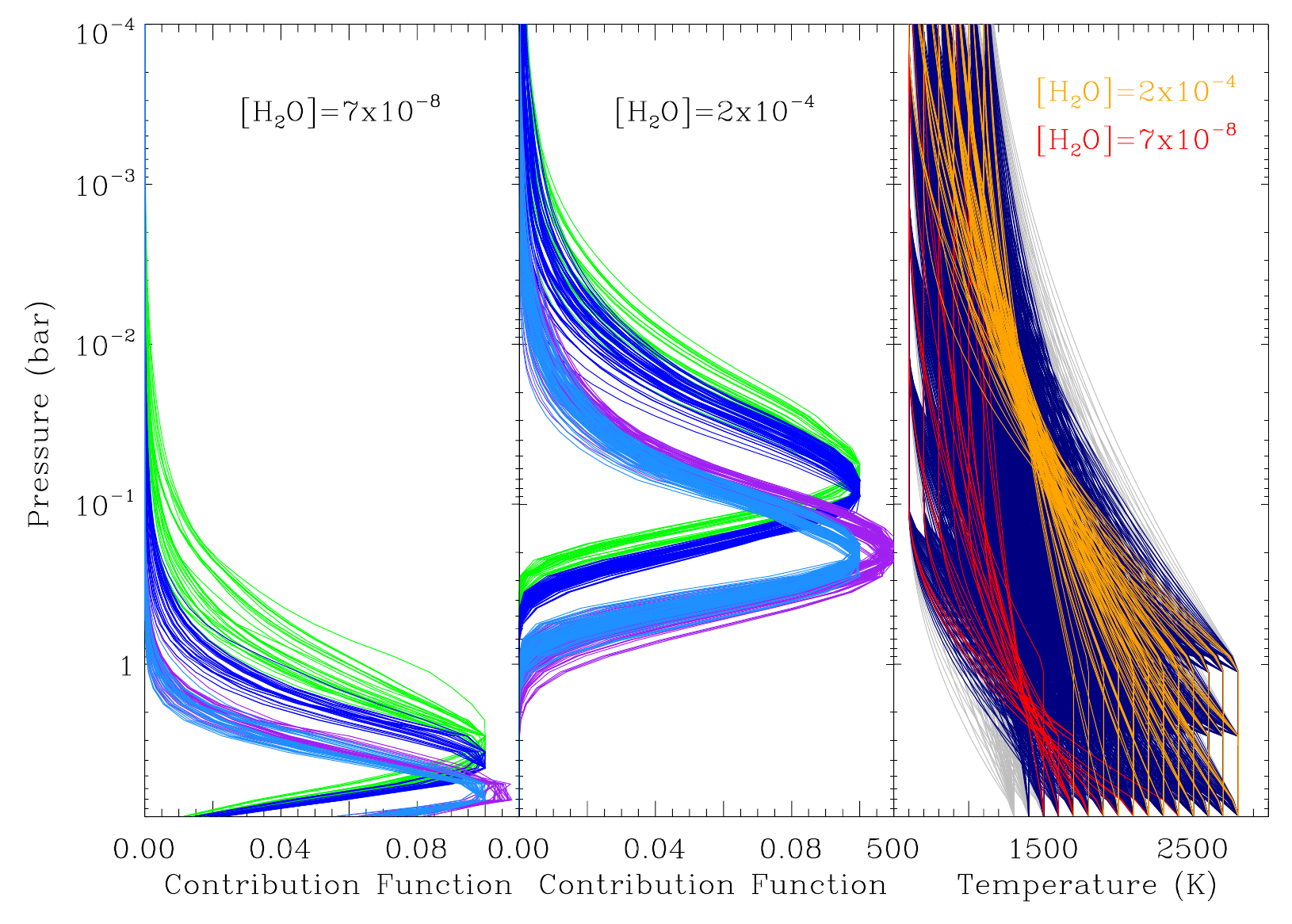}}
\caption{Right Panel: Most temperature profiles (blue) of the non-inversion models (grey) fit the data. The profiles 
separate into different subsets that depend on the model's water abundance, which dominates the opacity 
This effect is shown for 
models with [H$_2$O]=7$\times10^{-8}$ and [H$_2$O]=2$\times10^{-4}$. 
In each case, the contribution functions peak at the pressure level where the 
temperature is $\sim$1500 K. 
The water abundance
establishes the pressure level where most of the emission derives, as indicated by the contribution
models for the two water abundances. Contribution functions are shown for all the temperature 
profiles that fit the data at the Spitzer wavelengths of
3.6, 4.5, 5.8, and 8.0 $\mu$m, in light blue, purple, blue and green, 
respectively. Credit: Griffith {\it et al.} 2013.} 
\end{figure}

\subsection{Primary Transit Data} 

During primary transit the star's light drops 
by a factor equal to the ratio of the planet to star's effective areas.  
The depth of the resulting light curve at each wavelength, often termed ``absorption'', can be written \citep{Brown01} as: 
$$ 
A = {{\pi R_P^2}\over{\pi R_S^2}} ~+~
 \int_{R_P}^{\infty} 2 \pi R (1-Tr(R))dR / {\pi R_S^2}.  \eqno{(1)}
$$ 
Here R$_P$ is the radius of the planet at a specified pressure $P_0$ (e.g. 10 bars) where the planet is opaque; 
R$_S$ is the primary star's radius; and $Tr(R)$ is the atmospheric 
transmission at the particular wavelength of light through a chord that is a distance $R$ from the planet's center, referred to here as the impact radius (Fig. 5).  
The first term represents the occultation of 
the opaque part of the planet at pressures greater than $P_0$. For a Jupiter-sized body orbiting a solar-sized star, this term 
causes a $\sim$1\% drop in the light of the star. The second term represents the occultation of the 
planet's atmosphere, which causes a drop of $\sim$0.2\% or less, depending on the atmospheric scale height.  
Note that Equation 1 assumes that all the 
chords at an impact radius $R$ intersect the same atmosphere, although there are likely latitudinal 
and longitudinal variations in composition, temperature and cloud cover \citep{Cho03,Showman09,Fortney10}. 

The atmospheric transmission for a specified wavelength, $Tr(R)$, of light through a chord, $s$,  
that is a distance $R$ from the planet's center can be expressed as: 
$$
Tr(R) = {\rm exp}~ \left( -\int_s N(r(s))~\kappa_e(r(s)) ds \right)  \eqno{(2)}
$$
where $r(s)$ is the distance to the center of the planet at a point along 
the tangent, $N(r(s))$ is the atmospheric density at $s$, and $\kappa_e$ is
the extinction of all sources of opacities at the specified wavelength (Fig. 5). All scattered radiation is assumed
to leave the beam; thus $\kappa_e$ includes extinction due to both gas absorption and particle scattering. 
The optical depth is derived to the needed precision by dividing 
the tangential path into infinitesimal lengths, $ds$, and calculating the 
column abundance, $N(s)ds$, of the discrete bits of atmosphere, which when summed 
provide the total tangential column density, $N_t(R)$ at the impact radius $R$. 

\begin{figure}
\centering
%\rotatebox{0} {\includegraphics[type=eps,ext=.eps,read=.eps,width=2.5in] {figs/Primary-transit}}
\rotatebox{0} {\includegraphics[width=2.5in] {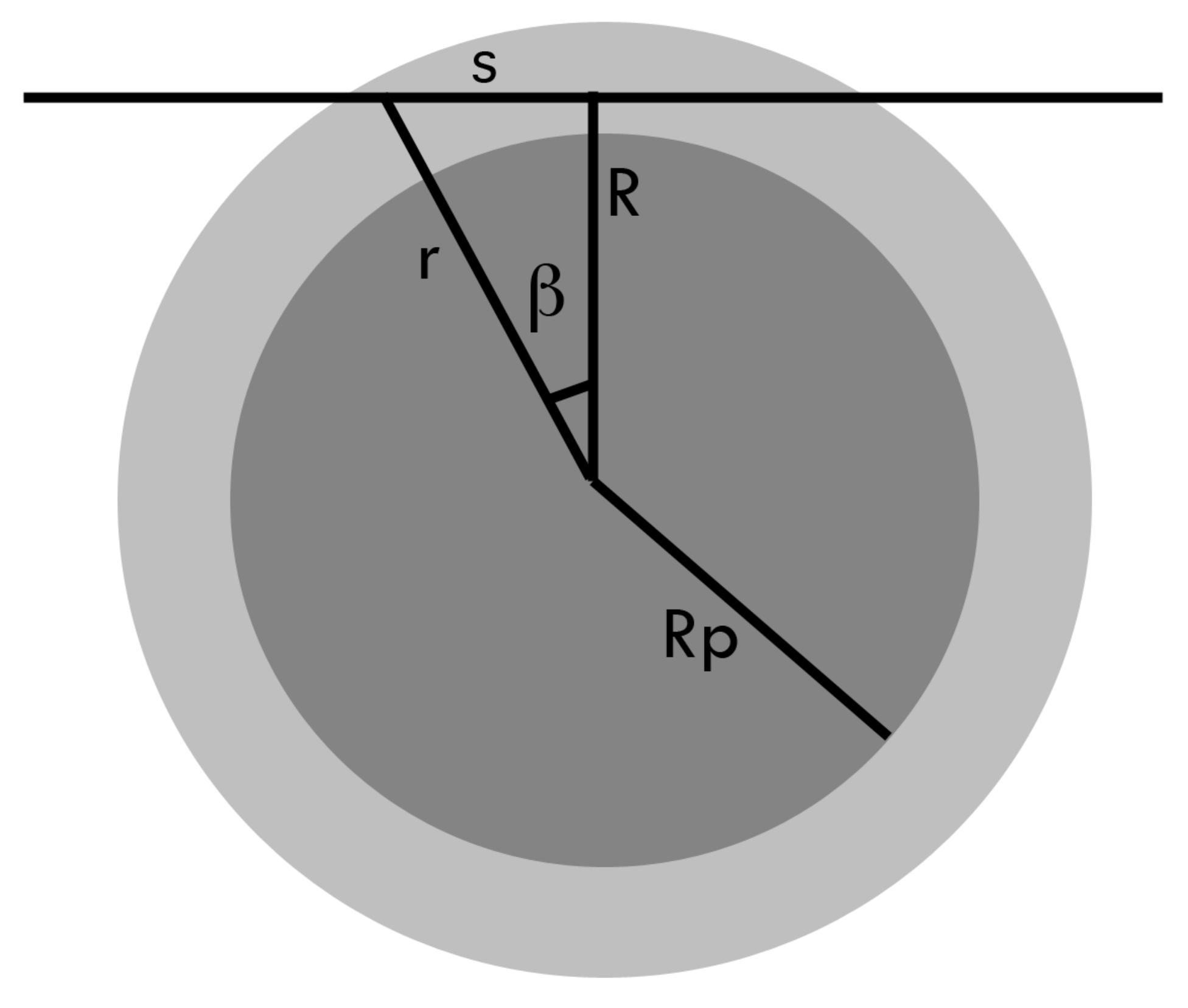}}
\caption{During primary transit, the star's light passes through the limb of the 
exoplanet. Shown is the path of one ray at an impact radius $R$. The ray traverses a 
total column density $N_T(R)$ of atmosphere, given by the density at each point 
along the chord integrated by $ds$, the infinitesimal distance. 
A reference radius, R$_P$, at a specific pressure is specified 
where the planet's atmosphere is opaque.  
} 
\end{figure}

As shown in past studies \citep{Tinetti10,Benneke12}, the 
analysis of primary transit data requires constraints 
on the temperature and radius at a specified pressure level.  
Before discussing the joint analysis of the primary 
transit and secondary eclipse data, we calculate the sensitivity 
of the derived mixing ratios on the planet's temperature and radius.

\section{Primary Transit Measurements}

\subsection{Approximate Eclipse Depth}
 
To develop an intuition of the nature of exoplanet transmission spectra, 
we work with an analytical approximation of the transmission of light through 
the planet's limb \citep{Chamberlain88}.  The extinction coefficient, $\kappa_e$, is assumed to equal 
that at the impact radius, i.e. $\kappa_e(s(r))=\kappa_e(R)  \forall s(r)$.  
The integral of the density along tangent line, $s$, an infinite distance in either direction, 
$$
N_t(R) = \int_{-\infty}^{\infty} N(r(s)) ds, 
$$ 
is approximated by assuming that the 
temperature along the chord is constant (Fig. 5).  
The value of $N(r)$ can 
then be expressed in terms of that
at the impact radius, $R$:  
$$
N(r) = N(R) e^{-(r-R)/H},	\eqno{(3)}
$$
where $H$ is the atmospheric scale height. 
The resulting integral is a modified Bessel function of the third kind, $K_1$, 
which has a series solution\footnote{
Change the dependent variable to $\beta$, the angle 
between $R$ and $r$ (Fig. 5). Then:  
$
r = R ~{\rm sec} \beta,
$ 
and
$
s = R ~{\rm tan} \beta,
$
and 
$
ds = R  ~{\rm sec^2} \beta  ~d \beta.
$
We find then that : 
$
N_t(R) = \int_{-\pi/2}^{\pi/2} N(R) e^{-R ({\rm sec} \beta-1)/H} ~ R ~{\rm sec^2} \beta ~d \beta,
$ 
the solution of which is a modified Bessel function of the third kind, $K_1$: 
$
N_t(R) = 2 N(R) ~R  ~e^{R/H}~K_1 (R/H) 
$ 
}.   
Keeping only the first term: 
$$
N_t(R) \approx N(R) ~(2 \pi R H)^{1/2}.  \eqno{(4)}
$$
This approximation is reasonably reliable because 
most of the contribution to the optical depth
derives from the closest point of the tangent line from the center of
the planet, i.e. the impact radius $R$. 
Correspondingly, the transmission of the atmosphere through a chord a distance $R$ away 
from the planet's center is: 
$$
Tr(R) = {\rm exp} (-N(R) ~(2 \pi R H)^{1/2}~\kappa_e).    \eqno{(5)}
$$ 

%\begin{figure}
%\centering
%\rotatebox{0} {\includegraphics[type=eps,ext=.eps,read=.eps,width=2.4in] {../figs/tangent-density-crop}}
%\caption{Integration of the tangent column density.} 
%\end{figure}

The transmission of the host star's light 
through the exoplanet's limb, i.e. the right-hand 
integral term in equation 1, is generally approximated as a sum over the impact radius. This sum  
consists mainly of terms that are either approximately zero or one, 
corresponding to regions where the chord traverses optically thin ($Tr(R)$=1) and thick  ($Tr(R)$=0)
slices of the limb, respectively. Basically as the impact radius increases,  moving outwards radially, the column density of the cord rapidly transitions from being opaque to optically thin, such that 
the region where $\delta Tr(R) / \delta P(R)$ peaks is where one is maximally sensitive. 
Information regarding the exoplanet's atmosphere can be derived 
only from the region in the atmosphere where the 
the transmission through the limb's chord most dramatically drops in value, 
e.g. from 5\% to 95\%.  

\subsection{Pressure region probed}

The pressure range probed by the transmission of light
through the limb of an atmosphere is 
$\sim$4 atmospheric scale heights. 
This is a general result, which we derive by 
designating $R_1$ as the impact radius where the
transmission through the chord is near one (e.g. $Tr_1=Tr(R_1)$=0.95) 
and $R_2$ the distance where the transmission is near zero (e.g. $Tr_2=Tr(R_2)$=0.05). 
The atmospheric region probed is then roughly 
$\Delta R = R_1-R_2$. 
Assume that the atmosphere 
is isothermal and homogenous within this pressure region; 
then, since  
$$
N(R_1) = N(R_2) e^{-(\Delta R)/H},  		\eqno{(6)}
$$
$\Delta R$ can be estimated by 
dividing ln($Tr_1$) by ln($Tr_2$) using Eq. 5: 
$$
\Delta R \approx H ~{\rm ln} [{\rm ln}(Tr_{2})/{\rm ln}(Tr_{1})]. 
$$ 
Adopting values of $Tr_{1}$=0.95 and $Tr_{2}$=0.05, 
we find that primary transit observations sample 
a region of roughly 4 scale heights ($4H$), 
i.e. a factor of $\sim$55 drop in pressure 
at each wavelength.   
Of course the atmosphere is not isothermal and  
homogeneous; this value is more indicative than 
exact. Therefore it is useful to compare this approximation 
to the results of a full model of 
of XO-2b's atmosphere, where the chord's column abundances, the 
temperature and pressure values, and their effects on 
the extinction coefficients are calculated precisely. The 
full calculation (Fig. 6) indicates that, at a wavelength of 1.56 $\mu$m, 
the primary transit of XO-2b probes a pressure region 
corresponding to a factor of 75 change in pressure, equivalent to $4.3 H$, 
i.e. close the the approximated value above. 
Given the small extent of the limb that is sampled, the opacity of the atmosphere 
acts primarily to determine the cutoff radius, $R_0$, where the atmosphere 
becomes optically thick.

\begin{figure}
\centering
\rotatebox{0} {\includegraphics[width=4.0in] {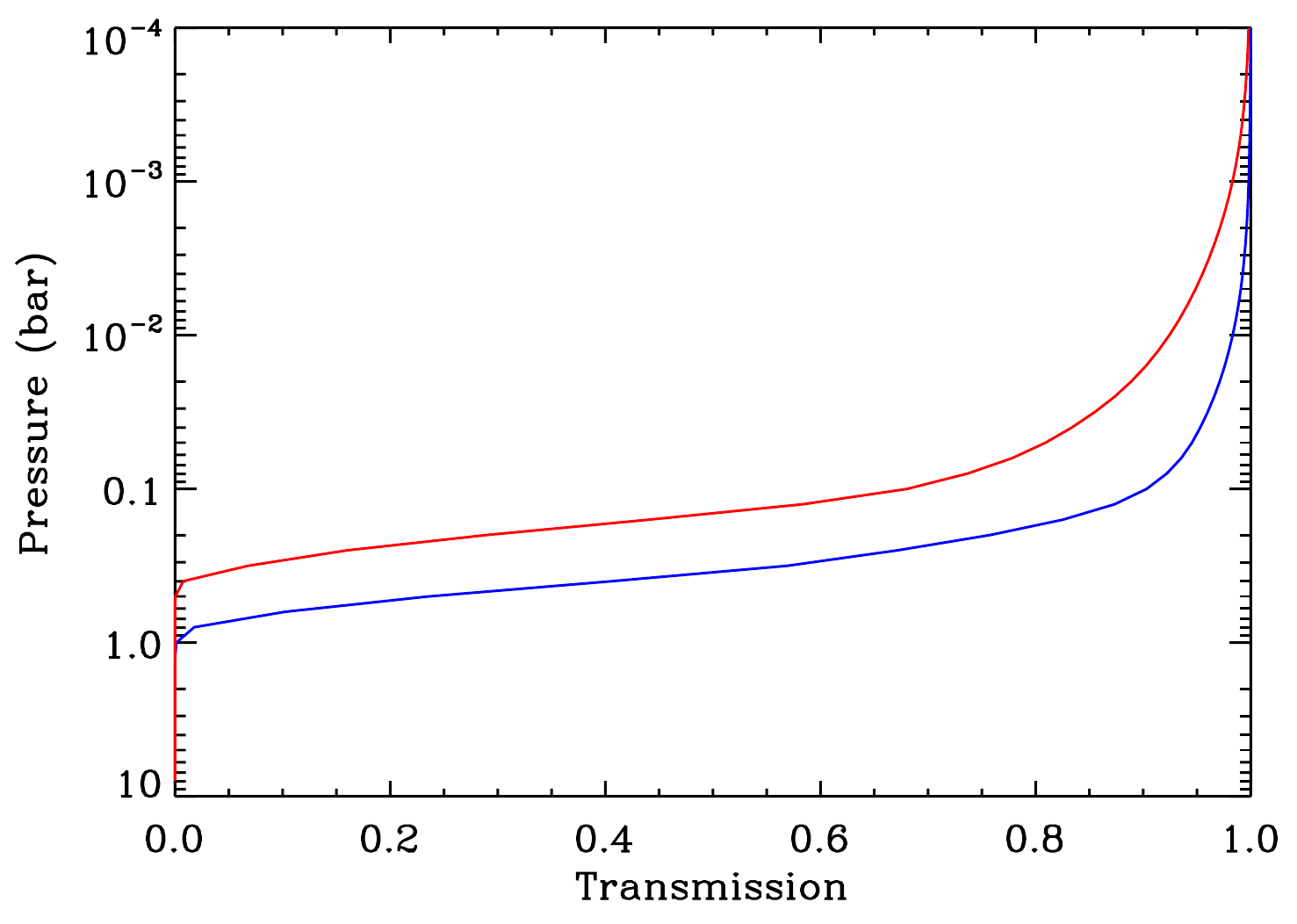}}
\caption{The transmission of 1.563 $\mu$m radiation through the 
limb of XO-2b as a function of the distance from the 
planet's center defined by the pressure at the impact radius. Shown are the transmission 
of a model atmosphere for a radius of 0.942 (blue) and 0.952 (red).}
\end{figure}

\subsection{Retrieval dependence on Radius}

As shown in Eq. 1, at a given wavelength the contribution of the 
atmosphere (right hand term) is established 
by the depth of the lightcurve, $A(\lambda)$,  (left hand term) 
and the assumed radius at a specified pressure, $R_p$ (middle term).  
The question then arises: how sensitive is the derivation of an exoplanet's 
composition to the assumed radius, $R_p$?  Knowledge of the detailed structure 
of the atmosphere is not needed to approximate this value. 
Consider the possibility that a given planet has a 
radius, $R_P$, at specified pressure (e.g. 10 bars) for 
which the uncertainty ranges from a large value, $R_L$, to a smaller value, $R_S$.
One can imagine two model atmospheres where for the $R_L$ model pressures are 
shifted upward by $\Delta R=R_L-R_S$ in comparison to the $R_S$ model. 
The ratio of the limb optical depths of the larger to smaller models at an impact radius $R$ is:
$$
{{\tau(R)_L}\over{\tau(R)_S}} = {{N(R)_L ~(2 \pi R H_L)^{1/2}~\kappa_L}\over{N(R)_S ~(2 \pi R H_S)^{1/2}~\kappa_S}}.   \eqno{(7)}
$$
At a specific impact radius, $R$, the integrated traverse column density of the 
larger radius planet is higher because the pressure is higher.
If the extinction coefficients are kept the same,  the increased column abundance of 
the higher radius planet will cause the optical depth of that model to be higher than that of 
the smaller model at $R$.   To achieve 
the same light curve, the extinction coefficient of the large radius model must decrease. 
Assume an isothermal atmosphere such that $H_L=H_S$ and assume a small difference between 
$R_L$ and $R_S$.  Approximate the density $N(R)_L$ in terms of 
$N(R)_S$ with Eq. 6; then the ratio of optical depths becomes:  
$$
{{\tau(R)_L}\over{\tau(R)_S}} = e^{(\Delta R/H)} ~ {{\kappa_L}\over{\kappa_S}}.   
$$
To match the observed light curve depth, the traverse optical depth at each impact radius 
must be equal for both the high and low radius models. Thus $\tau_L(R)=\tau_S(R)$, and  
$$
{{\kappa_L}\over{\kappa_S}} = e^{(-\Delta R/H)} 		\eqno{(8)}
$$
This expression indicates the sensitivity of the derived opacity 
to an uncertainty of $\Delta R$ in the assumed radius.  
For example, consider an uncertainty of 1\% in the planet's 
radius. This distance corresponds to roughly 4.6 atmospheric scale heights (H=146 km), 
assuming a jovian mass and radius, and a temperature 
(1000 K) typical of hot exoplanets. 
In this case, a 1\% uncertainty in radius of the planet indicates an optical depth 
uncertainty of a factor of 100 (Eq. 8).  

Constraints on gas abundances depend on whether the lines are in the weak or strong line limits.  In the 
weak line limit the curve of growth (i.e. the absorption) depends linearly on the abundance.  
In this case, if the radius is increased by 1\% and the optical depth at a given 
pressure level increases by 100, the gas abundances need to 
be decreased by 1/100 to bring the optical depth of the 
larger planet in agreement with that of the small planet.  Strong lines affect the 
absorption as the square root of the abundance; in this case the gas abundances 
need to be decreased by a factor of 1/10000.  In summary, mixing 
ratios derived from primary transit data are extremely sensitive to the 
uncertainty in the planet's radius.  

\subsection{Test with full model of XO-2b}   

The approximation above does not consider the dependence of the molecular cross section 
with pressure and temperature, which can be considerable.  To evaluate these effects, 
we calculate a full radiative transfer model of XO-2b's HST primary  
spectrum.   
This spectrum indicates primarily water absorption features \citep{Crouzet13}, allowing us to simplify this exercise and consider absorption of water only, although other gases could be included as well.  
Several models are calculated, each with the radius altered by roughly 1.05\%  (700 km or 0.01$R_J$) from an 
estimated value of 0.954 $R_J$ (Fig. 7). 
A change of 700 km is equivalent to 2.3 atmospheric 
scale heights.\footnote{The scale height of the exoplanet XO-2b is $\sim$300 km, larger than that the 1000 K hot Jupiter
considered above, because XO-2b is roughly half the size of Jupiter. This study assumes $M_{XO2b}$=0.566 $M_{Jup}$, \citep{Torres08}, although a newer study indicates a value of 0.62 $M_{Jup}$ \citep{Narita11}, the $\sim$10\% difference of which does not significantly affect this analysis. We assume a value of 71000 km for the radius of Jupiter. 
} 
For each model the gas abundance is recalculated to match the data (Fig. 7).  
The results of the radiative transfer models 
show that a 1\% change in the XO-2b's radius leads to an inferred abundance that differs from the original value by a factor of  $\sim$10 for mixing ratios of 10$^{-8}$ -- 10$^{-6}$ and by a factor of $\sim$100 for [H$_2$O]=10$^{-6}$ -- 10$^{-2}$, consistent with the progression from the weak to strong line limits.

\begin{figure}
\centering
\rotatebox{0} {\includegraphics[width=4.0in] {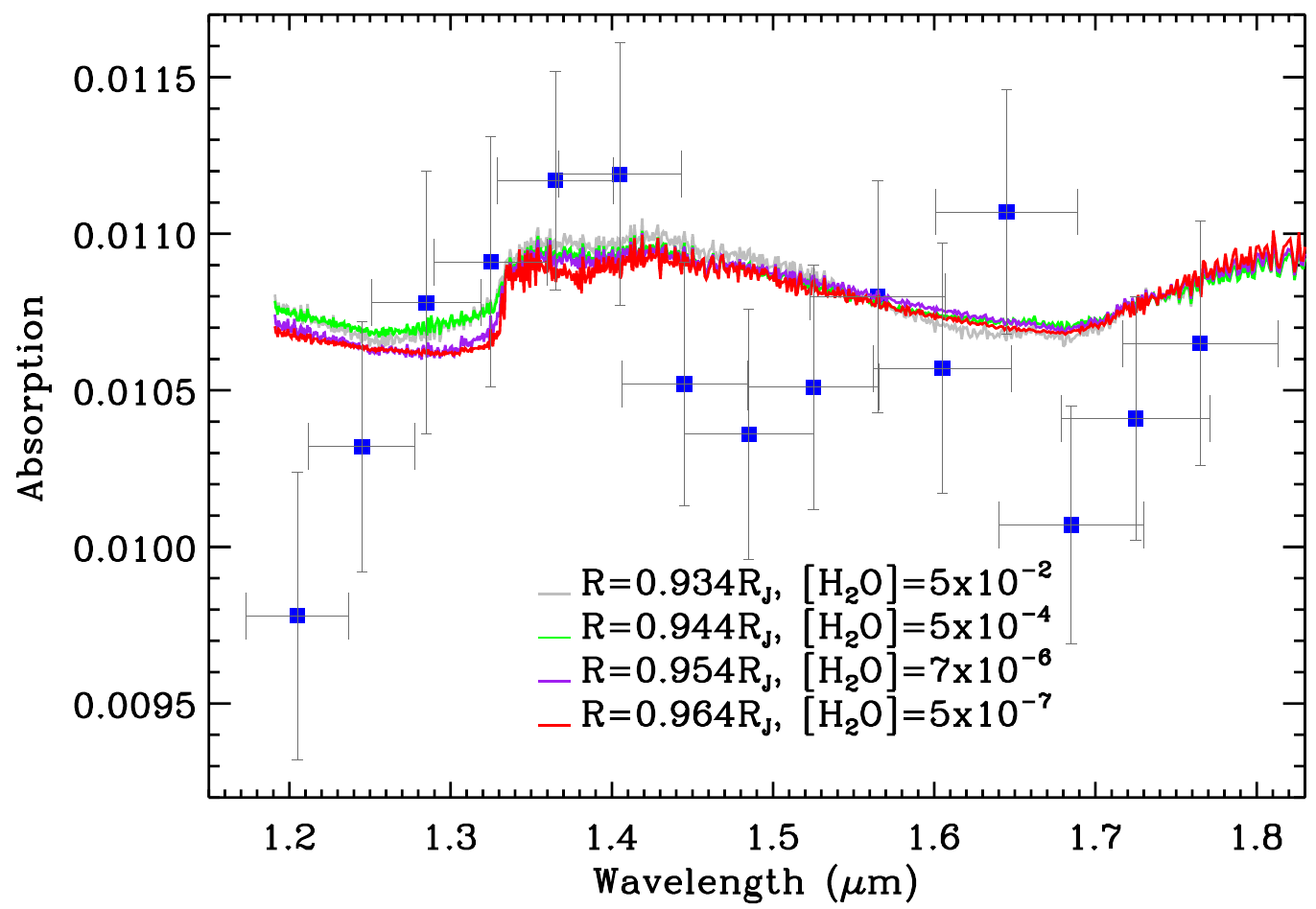}}
\caption{HST primary transit measurements of XO-2b \citep{Crouzet13} compared to
models calculated for radii that differ consecutively by 1.05\%. To produce 
similar spectra the 
water gas abundances were changed by factors of 14, 71 and 100, 
as the radius increases in value.  The increased factor in the mixing ratio results from the line regime, which progresses from the weak line limit to the strong line limit as the gas abundance increases.
(These models are not necessarily solutions, as the 
radius, composition and thermal profile must match both the emission 
and optical transmission data.) 
}
\end{figure}

Let us compare this calculation with what we would expect from our simple approximation (Eq. 8). First we need to 
determine if the absorption through the limb of the planet obeys the strong or weak line limit.  This can be 
accomplished by studying the dependence of the effective absorption (-ln$(Tr)/N_T(R)$) through the limb 
on the assumed water mixing ratio.\footnote{We work with an ``effective absorption'' because the extinction is calculated from k-coefficients.} The effective absorption at 1.56 $\mu$m for water mixing ratios of [H$_2$O]=10$^{-4}$ and 10$^{-5}$ (Fig. 8) indicate two different regimes, depending on the pressure of the 
impact radius. For values less than $\sim$10$^{-3}$ bars, the effective absorption of the [H$_2$O]=10$^{-4}$ and 10$^{-5}$ models differ by a factor of 10; here in the outer limb the absorption depends linearly on the mixing ratio. 
Deeper down, at pressures below 0.01 bar, the absorption differs by $\sqrt{10}$ $\sim$3 (Fig. 3). 
For the pressure level probed at 1.56 $\mu$m, $P_P\sim$0.5 bars (Fig. 1),  absorption obeys the strong line limit (Fig. 8). 
%Although the low abundance water model samples higher pressures,
%%, e.g. $\sim$0.8 bars at 1.56 $\mu$m wavelength, 
%where the absorption coefficient may exceed that at the levels
%%, 0.13 bars at 1.56 $\mu$m,  
%sampled by the high water content model, here this effect is not considerable. 
\begin{figure}
\centering
\rotatebox{0} {\includegraphics[width=4.8in] {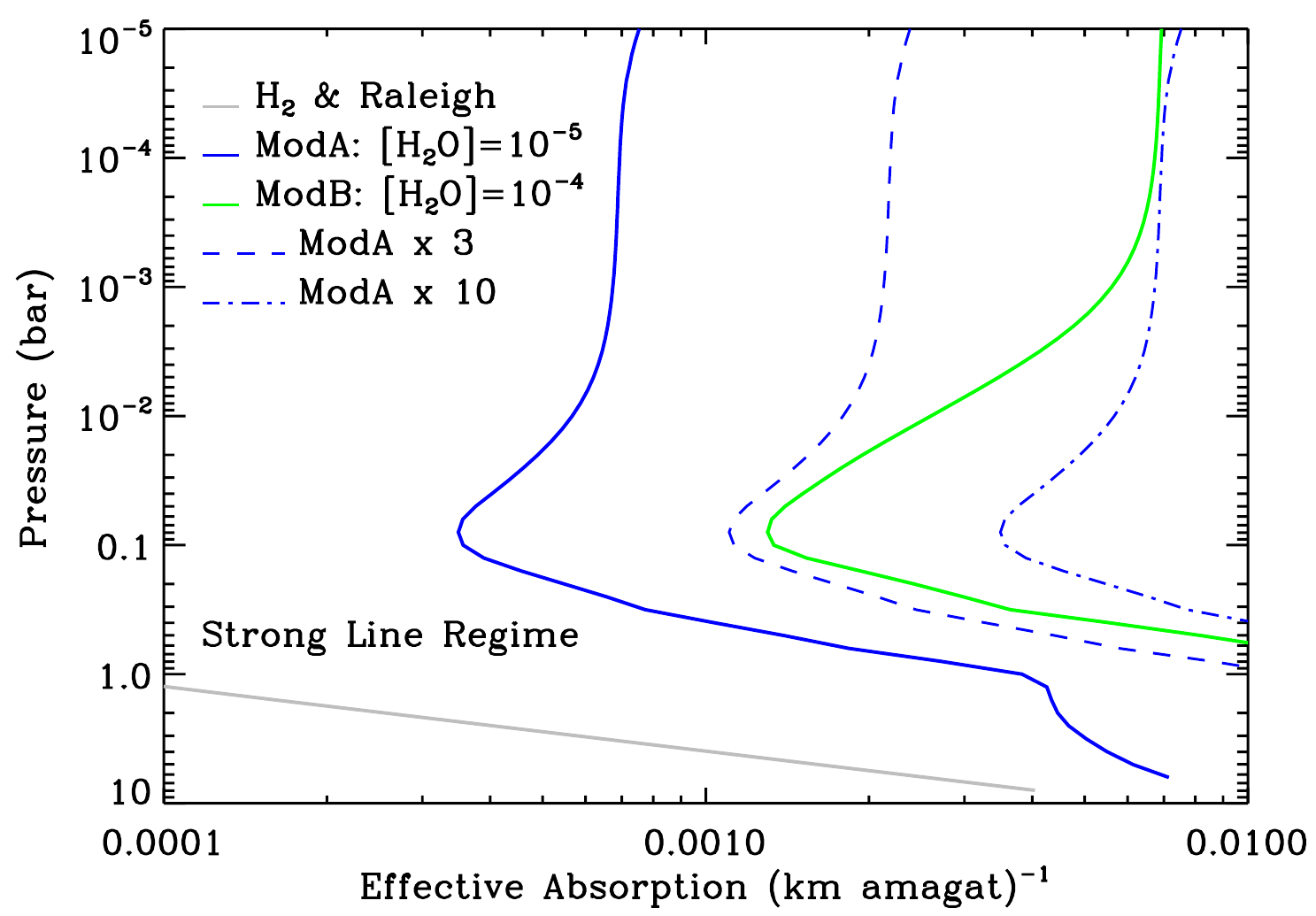}}
\caption{The effective absorption through the limb of XO-2b as a function of the impact radius for two model atmospheres of XO-2b (blue and green solid lines). These models are equivalent except that the green model has an order of magnitude less water than the blue model. Above a pressure of $\sim10^{-3}$ bars, the absorption coefficients of the green model resemble those of the blue model divided by 10; here absorption follows the weak line regime. Below 0.01 bar the green model absorption coefficients follow those of the blue model divided by the $\sqrt(10)=3$; here absorption follows the weak line limit.}
\end{figure}

Based on Eq. 8, an increase in XO-2b's radius by 1\%, the 
equivalent of 2.3 scale heights, requires that the optical depth be adjusted downward by a factor of 10 to produce the same spectrum.  Since the transmission spectrum probes levels below $\sim$10$^{-3}$ bars, absorption follows the strong line limit for water abundances of $10^{-5}$--$10^{-4}$ bars. Our approximation therefore indicates that the water mixing ratios greater than $10^{-5}$ must be decreased by a factor of 1/100 to fit the data, if the radius is increased by 1\% .  This result agrees with that of the full model (Fig. 7) and indicates that
%that depending on the line regime and the scale height, uncertainties in the radius affect uncertainties in the derived abundances. In all cases the (of for most planets studied currently) 
the certainty in the derivation of composition of a planet's atmosphere from primary transit data depends strongly on the precision with which the radius is known. For many planets this value is not well known; 
for example the radius of one of the most extensively 
measured exoplanet, HD189458b, is uncertain by 6\% \citep{Torres08}.

%Nonetheless, this exercise indicates the need for establishing the radius of the planet as a function of pressure. 

\subsection{Temperature dependence on primary transit retrievals}

The derived composition of an exoplanet is also sensitive 
to the assumed temperature structure through its effects on 
the atmospheric density structure and 
on the extinction coefficient.  A cooler atmosphere is more 
compressed (smaller scale height) and exhibits a smaller  
light curve depth than does an otherwise identical warmer 
atmosphere. As a result, uncertainties in the 
temperature structure of an atmosphere lead to uncertainties in 
the abundance.  

This sensitivity can be estimated roughly by calculating the 
impact radius where the traverse transmission $Tr= 0.5$ for a hot and cool planet; 
that is the radius, $R_0$, where the atmosphere 
becomes optically thick. 
We consider, as an example, the effect of a 300 K uncertainty 
in the temperature profile, which is roughly that derived from studies of 
the dayside emission of HD 189733b \citep{Madhu09}.  
At the wavelength 0.45 $\mu$m,  
where the atmosphere's opacity is dominated by 
Rayleigh scattering, the extinction coefficient is 
$\kappa_e=2.4 \times10^{-27}$ m$^2$/molecule.   At this wavelength 
and assuming the ideal gas law and Eq. 5, a hot isothermal Jupiter 
exoplanet with $T=1000 K$, $M=M_J$, $R_r=R_J$, and $H$=146 km, 
has a transverse chord transmission of 0.5 at 0.049 bar. This pressure level 
lies 776 km above our 10 bar reference level, $R_p$. If  the temperature is  
increased everywhere by 300 K, i.e. to 1300 K, the scale height becomes 190 km. 
We find that the $Tr$=0.5 chord transmission occurs at 0.056 bar, 
which lies 983 km above the 10 bar  level. The radius of the planet increased by 207 km. 
In order to achieve a transmission of 0.5 at the same radius, 
the optical depth of the hotter model must be 
decreased to one forth the original value (Eq. 8) 
so that the radius decreased by 207 km. 
This estimate is of course rough, as we have assumed an 
isothermal atmosphere and considered only one absorption 
coefficient. 

To explore these effects further, a full radiative transfer model 
of XO-2b is compared to an analytical approximation, like that above.  
Considering first the analytical approach, 
the pressure level of the $Tr$=0.5 transmission at 0.45 $\mu$m wavelength
for XO-2b is approximated (using Eq. 5)to be 0.04 bar, assuming a 
ballpark temperature of 1000 K and the atmospheric scale height of 300 km of XO-2b. 
A 300 K warmer planet has a $Tr$=0.5 transmission at 0.045 bar and the 
distance between the 10 bar level and the $Tr$=0.5 transmission 
level increased by 351 km (Eq. 8).  
At near-IR wavelengths, given the probed temperatures and pressures, 
the effective extinction coefficients range 
from values similar to that at 0.45 $\mu$m to values in the more transparent regions 
that are roughly 10 times less, e.g. 
$\kappa_e=2.4 \times10^{-28}$ m$^2$/molecule, at a spectral 
resolution of R$\sim$50.  Assuming this absorption, the transverse 
chord has a $Tr$=0.5 transmission at 0.46 bar (Eq. 5). A 300 K warmer planet 
has a $Tr$=0.5 transmission at 0.40 bar (Eq. 8), and the distance between the 10 bar 
level and the $Tr$=0.5 transmission level increased by 187 km. 
If primary transit data exists at 0.45 $\mu$m,  
$R_p$ is decreased so that the hot model fits the data, 
in which case the hot model's radius is too low to fit the near-IR 
data by 164 km = 351--187. 
In order to interpret both the optical and more transparent IR data 
the optical depth of the hotter model must be 
increased to half the original value (Eq. 8).  Therefore, 
since the strong line limit applies, the mixing ratio must 
increase by a factor of 2--4, in the spectral regions that 
are most transparent depending on whether one is in the 
strong or weak line limit.  

Considering now a full model atmosphere, 
the cooler model is assumed to have a 
temperature of 600 K above the 0.1 bar 
tropopause pressure and a 2500 K
%2700 K   in fact this was the temperature but it is not shown in figure 4
temperature below 2.8 bar. The warmer model is everywhere hotter by 300 K. 
We find that a hotter model, where temperatures are everywhere hotter by 300 K, 
indicates gas abundances that are 1.5 to 3 times greater 
%(in the more optically thick spectral regions) 
compared the abundances indicated 
in the cooler model. This calculation agrees, within a factor of a few, with the 
analytical model. Thus uncertainties in the temperature profile affect 
the precision of the derived composition, however less significantly 
than do the uncertainties in the radius.  

\section{Resolving the Degeneracies} 

\subsection{Determination of the radius}

The study above indicates that derivations of composition
of hot jovian exoplanets, while sensitive to the 
inferred temperature profile, are extremely contingent on
the assumed planetary radius as a function of pressure. 
Errors in the derived radius arise from two sources: 1) 
uncertainties in the measured radius of the roughly unity 
optical depth, e.g. \citep{Torres08}; and 
2) uncertainties in the assignment of the pressure level associated
with the retrieved radius at unity optical depth. These uncertainties, 
on the level of a few percent of the planet's radius (i.e. over 1000 km), 
exceed the range of radii that are currently considered in most 
inversion studies of exoplanetary spectra.   

High spectral resolution observations of Solar System planets resolve  
the pressure of an absorption feature through the line widths of absorption features, which
are established by pressure broadening in the troposphere. 
Yet this technique does not work as well for hot jupiter atmospheres, which 
are hot enough that doppler broadening establishes the line widths 
over most of the observable atmosphere.\footnote{
The doppler broadening full width half maximum (FWHM) of a 1000 K atmosphere exceeds the pressure broadening FWHM (assuming a standard STP value of 0.05 cm-1) at all wavelengths below 4 $\mu$m and all pressures less than 1 bar.  Since the pressure broadening FWHM depends linearly with pressure, Doppler broadening dominates over pressure broadening for hot Jupiters except at high wavelengths that probe deep levels.}
In addition, the faint nature of exoplanets hinders
high resolution spectroscopy. 

A more appropriate method for eliminating degenerate solutions 
to the primary transit data is to measure the radius at a wavelength where
the atmosphere's opacity, and thus the probed pressure level, is known \citep{Etangs08,Benneke12}.  
For the atmospheres of hot jovian exoplanets the region between 0.3--0.5 $\mu$m 
is defined mainly by Rayleigh scattering, as long as the atmosphere is cloudless. 
If the atmosphere is cloudy, such measurements define the upper 
limit of the radius at a specific pressure, since clouds  add an unknown opacity source. 
For the brightest exoplanets, radii can be constrained 
using U ($\lambda \sim$ 0.35 $\mu$m) and B ($\lambda \sim$ 0.45 $\mu$m) 
photometry with relatively small telescopes (e.g. 1.6 m) 
equipped with a sensitive CCD and a high cadence readout \citep{Dittmann09a,Turner13,Teske13b}. 
In addition star spots, which can considerably  
affect optical measurements, can be characterized  
through broad-band optical and near-IR photometry \citep{Ballerini12}.
Larger ground-based telescopes are able to obtain 
optical spectra, which have the advantage of defining the slope of the 
opacity and thus potentially the distinctive Rayleigh signature, which
indicates gas scattering as opposed to that of clouds or haze. 
Small telescope optical observations of exoplanets 
outside of absorption features such as Na and K probe 
spectral regions where the primary opacity of the atmosphere, 
if cloud free, is known. Observations at these wavelengths constrain the radius pertaining
to a specific pressure level, thereby  
enabling interpretations of IR primary transit 
measurements from HST, Spitzer and ground-based platforms.  
This leverage can be seen in the analysis of primary transit and secondary eclipse data  
of  XO-2b \citep{Griffith13}.

\begin{figure}
\centering
\rotatebox{0} {\includegraphics[width=4.5in] {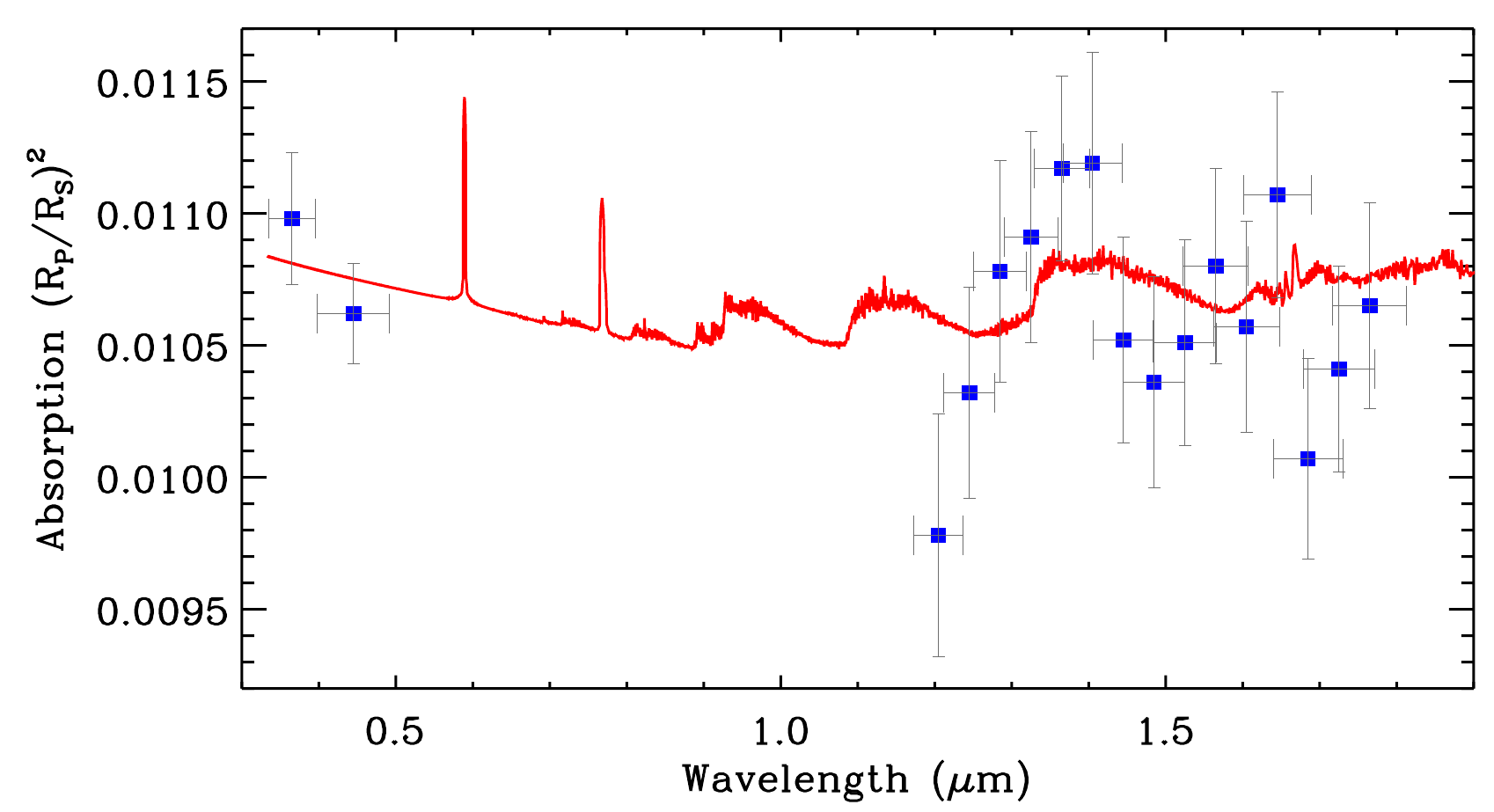}}
\caption{Primary transit near-IR spectrum \citep{Crouzet13} and optical photometry measurements
\citep{Griffith13} of XO-2b (blue squares) are compared to a model spectrum (red).}
\end{figure}

\subsection{Joint analysis of primary and secondary transits}

The first optical photometry of XO-2b, conducted in the 
broad Sloan z band (effectively $\sim$0.83--1.00 $\mu$m), 
yielded a measurement of the planet's density and  radius of 
0.996 $R_J$,  corresponding to the level where the tangent chord is optically 
thick over the wavelengths sampled \citep{Fernandez09}.   It is 
difficult to associate this radius to a pressure level because the 
band covers spectral features due to K \citep{Sing12}, Na and water.  
Griffith et al. (2013) therefore recorded photometry of XO-2b at 
the 1.55 meter Kuiper Telescope on Mt. Bigelow 
in the U-band and B-bands, outside the effects of atomic and molecular 
spectral features, where Rayleigh scattering establishes the opacity, if cloud-free.  
Data were obtained between January 2012 and February 2013 
over the course of 4 nights, yielding 3 measurements 
of the U band (0.303--0.417 $\mu$m) and 2 of the B band (0.33--0.55 $\mu$m) \citep{Griffith13}. 
Here we consider  
the data from 9 December 2012, the one night in which 
both U and B measurements were simultaneously recorded (Fig. 9), and discuss the implications
of the other measurements below.  

The observations are analyzed to determine the planetary radius at 10 bar, which depends on the 
thermal profile. The range of values that fit the U and B data span about 
1\% of the planetary radius (Fig. 9), causing, according to our approximations, an uncertainty of a factor of 100 in the derived
abundance assuming a fixed temperature profile.  However since the temperature is not constrained
there is a larger range of solutions. 
The temperature profiles that fit the secondary eclipse data range by 1000 K, 
suggesting an uncertainty of roughly an order of magnitude in the derived gas abundances
from the primary eclipse due to the temperature range for a given assumed radius. 
But then the radius that fits the data depends on the temperature profile, 
as also demonstrated in models of test hot Jupiter spectra \citep{Barstow13}. 
These considerations indicate the difficulty in deriving constraints from 
either primary or secondary eclipse data alone, and point to 
a coupled analysis of both emission and transmission data. 
In fact, such an analysis, described below, shows that  
coupled temperature and abundance profiles derived from the 
secondary eclipse data can form quite a separate solution set 
from that of the primary transit data, which overlap in 
a small region of phase space.

\begin{figure}
\centering
\rotatebox{0} {\includegraphics[width=4.0in] {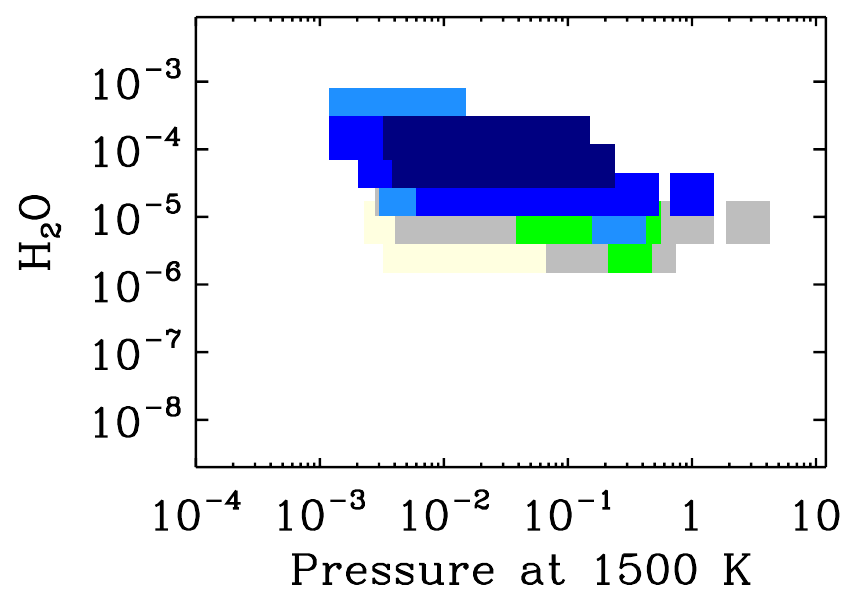}}
\caption{The subset of models that fit both the emission and transmission measurements
of XO-2b. The blue squares indicate the best fits to the primary transit spectrum
\citep{Griffith13}.}
\end{figure}

To interpret both primary transit and secondary eclipse data we begin by assuming cloudless conditions. 
First the radius, R$_P$,  is determined from the U and B band light curves 
separately for each thermal profile considered in the analysis of the secondary 
eclipse measurements of \citep{Machalek09}.
Analyses of the secondary eclipse data provide a solution set of coupled thermal profiles 
and composition (e.g. Fig. 3).  These models are fed into a calculation of the primary transit light curves 
depths, which are compared to the near-IR transmission data \citep{Crouzet13} to determine the models 
that matches both secondary and primary eclipse measurements (Figs. 1 \& 9). 
The solution set
is calculated for both the largest and smallest radii to derive models that interpret all 
the data within errors. This analysis yields significantly 
stronger constraints on the water composition of XO-2b than does the interpretation 
of one type of data alone, as shown for water in Fig. 10.   

The differences between the terminator and the dayside hemisphere compositions 
(sampled during primary transit and secondary eclipse respectively) depends 
on the species and pressure level probed.  
Water, which likely dominates XO-2b's spectrum, and CO are expected to 
have constant abundances over the pressure levels probed, both vertically 
and horizontally, as they are not
affected by photochemistry below $\sim$10$^{-6}$ bars. Methane and carbon dioxide, more strongly  
affected by photochemistry, 
are predicted to have variable vertical profiles \citep{Moses11}, which, with low spectral resolution observations, are measured as weighted average abundances over the small pressure regions probed. 
These species also vary in abundance between the terminator to dayside atmosphere depending on the insolation and vertical mixing.   
For example, predictions for HD 189733b lead to the hemispherical abundances that differ by factors of $\sim$4 and $<$2, for methane and carbon dioxide, respectively \citep{Moses11}. 
Jovian exoplanets, which are mostly locked in synchronous rotation about their host stars,
also display different temperature profiles (Fig. 2) at the terminator 
compared to the dayside atmosphere \citep{Harrington06,Knutson07,Crossfield10,Knutson12,Lewis13}. 
Models of the thermal profiles of the 
close analog system HD 189458b indicate dayside temperatures 100-200 K hotter 
than those at the terminator \citep{Showman09}, which, based on our estimates, leads to an overestimate in the gas abundances derived by the primary transit data by a factor of 1-2. 
%These studies indicate that a thorough analysis of transit data, therefore, would include full phase measurements of the planet's orbit to constrain the hemispherical variations in temperature and composition. 

Clouds have different effects on primary eclipse data depending
on the cloud particle sizes. One can envision two possible scenarios.  
The clouds could be of roughly uniform optical thickness from optical to IR wavelength 
if the cloud particle size exceeds 10$\mu$m. An example is the water clouds on Earth.  
Alternatively, clouds could be more optically thick at shorter wavelengths rather than at longer wavelengths
if they consist of sub-micron sized particles. Such particles describe the photochemical
haze on Titan \citep{Tomasko08a}. The presence of the former large particle clouds 
would increase the exoplanet's radius at wavelengths that probe
down to the cloud deck.  The model of the primary transit spectrum would
be flattened, contrary to what we observe.    The presence of the latter 
small particle clouds would increase the light curve depth at small wavelengths, 
with no effect on the near-IR spectrum.  In this case, the inferred gas abundances
would be larger than those determined for a cloudless atmosphere.  The entire
solution set would shift upward, and the abundance of 
water could be higher than that expected for a solar metallicity atmosphere
in thermochemical equilibrium.  Since the existence of such a cloud cannot
be discounted in this case, the measurement of XO-2b radius is in fact an upper 
limit, since the presence of a cloud would indicate a smaller radius. 
The upper limit of XO-2b's radius allows one to constrain the lower limit
of the gas abundances. 

The additional U and B measurements are particularly interesting because they
indicate the effects of processes that have not been considered in the analysis
above.    Specifically the measurement in the B band on 28 October 2012 
disagrees with that from 9 December 2012 at the three sigma level \citep{Griffith13}. Regarding the 
two remaining U values, the radius derived from the 5 Jan 2012 data is identical to 
that of 9 December 2012, while the 23 Feb 2013 value is higher by roughly 1 sigma \citep{Griffith13}.  
A more extensive discussion of the errors and their treatment is given in \citep{Griffith13}.
The spurious U value may result either from the presence of non-uniform clouds 
in XO-2b's atmosphere or from sunspots on the host star.  
Unaccounted for systematic errors in the data could also play a role.  
The higher radius derived from the outlying B and U measurements allow for 
a significantly larger solution set in the analysis of the secondary and primary measurements \citep{Griffith13}. 
In addition, they introduce the possibility that observations taken at different times, 
and therefore potentially different cloud or sunspot conditions, cannot be interpreted together. 
%Comparison of the flux of the host star to another reference star (e.g. the host's binary companion) 
%for each of the 3 nights of U measurements 
%did not shed any light on the cause of the large radius determined on 2013 Feb 23 -- 
%whether due to e.g. clouds or star spots -- 
%because the fluxes could not be determined with sufficient accuracy \citep{Griffith13}.  
The cause 
of the the variability in the measured flux can be investigated by optical  
spectra, which will yield the slope of the primary eclipse spectrum and thus whether
due to gas or cloud scattering.  Simultaneous photometric measurements from a 
number of telescopes will further illuminate the systematic errors.

\section{Summary}

The analysis of a sparse data set discussed above indicates 
that future prospects are  promising: the primary transit 
and secondary eclipse data considered together can, with greater spectral coverage 
and sampling, lead to strong constraints on the characteristics of exoplanet 
atmospheres.  However this exercise also indicates a lack of 
vital information that is needed to constrain the exoplanet 
characteristics. The rigor of the approach outlined here would be
improved with full phase observations which illuminates disk variations
in the temperature and composition profiles, (e.g. Knutson {\it et al.} 2012, Lewis {\it et al.} 2013). The exoplanet's atmosphere was assumed,
for lack of information, to be cloudless, although in fact, the variability in the 
measurements potentially points to variable cloud cover. As indicated 
by measurements of brown dwarfs(e.g. Kirkpatrick {\it et al.} 2005) and the directly measured 
HR8799 planets (e.g. Marois {\it et al.} 2008, Marois {\it et al.} 2010) observations of cloud effects depend
on the gravity, effective temperature and thus the 
sedimentation efficiency of the planet
\citep{Marley12}. In addition, we assumed no temperature 
inversion (i.e. as in a stratosphere) and
no non-LTE emission, both of which can effect 
specific features.  Observations in the L-band of
HD 189733b indicate 
non-LTE emission \citep{Swain10,Waldmann12}, 
and HD 209458b's spectrum indicates the 
presence of a stratosphere \citep{Burrows07,Madhu09}.
The interpretation of transiting planet measurements 
are also potentially affected by star spots, which are indicated by 
measurements of the host star HD 189733, (e.g. Pont {\it et al.} 2007, MillerRicci {\it et al.} 2008). 
In addition, many analyses of exoplanetary data,
including the primary transit measurements discussed here, do
not fit all of the data to within the error bars (Fig. 9).   

Smaller and cooler planets more similar to Earth present 
more challenges. As poor emitters, they 
cannot presently be measured with secondary 
eclipse observations.  One cannot assume
the composition of the primary constituent, and thus the 
mean molecular weight as we did here.   Also, 
the presence of a surface adds an additional variable to be 
investigated. As explored by models of Benneke and Seager (2012), 
these additional challenges can be addressed by 
recording high signal to noise spectra over a 
broad wavelength region (0.5--5 $\mu$m), possible for some planets with
upcoming James Web Space Telescope (JWST) \citep{Benneke12}. 
The lack of secondary eclipse spectra is treated  
by making assumptions regarding a radiative-convective thermal profile, based on 
albedo assumptions \citep{Benneke12}.   The mean molecular weight 
is decoupled from the derivation of other parameters 
by recording both weak and strong lines of the same 
constituent, the radius and Rayleigh slope from 
high S/N data extending from 0.4--0.7 $\mu$m, and   
the wavelength shift in the Rayleigh slope \citep{Benneke12}. 
Detailed studies of smaller and cooler planets 
may be best approached with large mirror
investigations, e.g. with the JWST.   

However, the combined analysis of primary and secondary eclipse data 
and the exploration of the degeneracy of 
solutions explored here indicates the 
need to better understand exoplanet data,  
the properties of host stars, and the 
range of assumptions inherent in any model. 
This task requires observations from ground 
and space of many extrasolar systems.  
A small space telescope survey of exoplanets, where   
each observation covers 
large wavelength IR region with redundant molecular features
with sufficient resolution to identify broad rotation-vibrational features, 
would constrain a significant number of basic parameters of hundreds of systems.  
Multiple observations of planetary systems from 
ground-based telescopes can explore 
time-variable processes such as clouds and star spots. 
In addition, there
is a need to target exoplanets of particular interest
with large telescopes, which perforce
will not have the time to sample a large number 
of planetary systems. 
This combined approach of survey telescopes
and large mirror telescopes in space, will provide 
detailed information on particular systems, and 
general information on a large number of systems, 
as needed to understand planet formation explore
and the variety of planets and planetary systems. 

\bibliographystyle{apj}
%\bibliography{cgexo}
\bibliography{coexo.bib}

\end{document}